\documentclass{llncs}

\usepackage{graphicx}
\usepackage{url}
\usepackage{color}    
\usepackage[ruled]{algorithm2e}
\usepackage{fancyvrb}  
\usepackage{hyperref}

\newcommand{\mysubscript}[1]{\raisebox{-0.34ex}{\scriptsize#1}}

\title{A Metapolicy Framework for Enhancing Domain Expressiveness on the Internet}
\author{Gaurav Varshney and Pawel Szalachowski}
\institute{SUTD, Singapore}
\date{}

\begin{document}
\maketitle

\begin{abstract}
Domain Name System (DNS) domains became Internet-level identifiers for
entities (like companies, organizations, or individuals) hosting services
and sharing resources over the Internet.  Domains can specify a set of
security policies (such as, email and trust security policies) that should
be followed by clients while accessing the resources or services represented
by them. Unfortunately, in the current Internet, the policy specification and
enforcement are dispersed, non-comprehensive, insecure, and difficult to
manage.

In this paper, we present a comprehensive and secure metapolicy framework for
enhancing the domain expressiveness on the Internet. The proposed framework
allows the domain owners to specify, manage, and publish their domain-level
security policies over the existing DNS infrastructure. The framework also
utilizes the existing trust infrastructures (i.e., TLS and DNSSEC) for providing
security. By reusing the existing infrastructures, our framework requires
minimal changes and requirements for adoption.  We also discuss the initial
results of the measurements performed to evaluate what fraction of the
current Internet can get benefits from deploying our framework. Moreover,
overheads of deploying the proposed framework have been quantified and
discussed. 

\end{abstract}


\section{Introduction}
\label{sec:intro}
Domain names are a de facto standard way to identify computers, networks,
services and other resources on the Internet. Domain security policies provide a
way through which domain owners can specify the restrictions or rules that
should be followed while accessing the computers, services or the resources
represented by their domain names. 

Currently, most of the domain security policies are either specified
individually and published using the DNS infrastructure (e.g.,
SPF~\cite{rfc7208}, DKIM~\cite{rfc6376}, DMARC~\cite{rfc7489} --- see
\autoref{sec:pre:security_policies}), or are specified at the domain web servers
and communicated to \textit{policy agents}~\footnote{A policy agent is a
software component that processes and enforces policies. It can be implemented
within a user agent (such as a browser) or within a server software that
supports a given policy.} via dedicated HTTP headers (e.g., HSTS~\cite{rfc6797}
or HPKP~\cite{rfc7469} --- see \autoref{sec:pre:security_policies}). Finally, the
obtained security policies are enforced by policy agents.

In some cases, the enforcement of security policies is not automated and requires
user's involvement (i.e., users are making policy decisions).  One example of such a case
is accepting or denying a secure connection to a domain that presented an expired
certificate.  However, most of the security policies are standardized and governed
by software vendors and Internet communities, and domains cannot influence this
process and have to just follow these standards for specification of their
security policies.   

The current mechanisms of security policy specification and enforcement are
unsatisfactory and the future Internet requires a higher level of domain
expressiveness for the following reasons.

\begin{enumerate}
\item Users are not proficient enough to make security decisions when a policy
    agent requires that~\cite{egelman2008you}. In the previous studies, it was
        observed that most of the users do not even notice the browser security
        indicators (like padlock icons), or they ignore warnings displayed to
        them by browsers and just
        \textit{clickthrough}~\cite{egelman2008you,kranch2015upgrading}. 
\item For scalability reasons, software vendors and the Internet community can
    only introduce global and generic policies without focusing on
        domain-specific policies. Obviously, global policies might not fit all
        domains as domains have different resources, services, and business
        models.  One concrete example is a non-security-critical website (like a
        news or an informational website) that mostly displays a read-only
        content to its visitors and makes profits on ads. In such a case, the
        website may want to relax its security policies and display the content
        (and ads) to visitors, even if some security properties are not met
        (e.g., the website's certificate is expired). On the other hand, an
        e-banking website may need a stricter security policy that must generate
        an error and does not let its users interact with the website in a case
        of certificate errors.  Domains are usually more aware of their security
        requirements and therefore they are the right candidates for
        policymakers.  Unfortunately, the current policy specifications barely
        consider domain-specific requirements as of today. 
\item Another consequence of policies implemented by software vendors is that
    these policies may be inconsistently enforced by different software
        implementations, especially, when a policy specification leaves some
        choices to developers.  For instance, if browsers do not implement
        policy enforcement uniformly, it may cause a situation where users can
        switch from one browser to another in order to overcome a generated
        policy error (actually, such a behavior has been observed in the past).
        Hence any new framework of security policy specification could benefit
        from providing a way through which security policies can be specified
        and managed by domains with a relatively less involvement of software
        vendors, user agents, or even the users. 
\item Downgrade attacks, like stripping of policy headers, is another problem.
    Policy headers can be manipulated by a Man-in-the-Middle (MITM) adversary,
        or at client-ends via modified implementation (like malicious browser
        extensions). Such a stripping of headers may lead to downgrade attacks,
        as an adversary can pretend to a client that the contacted domain does
        not deploy the given enhancement or policy. Third party extensions such
        as \textit{Modify headers for Google Chrome}~\cite{ModifyHeader} can be
        used to modify or strip off HTTP headers making it easier to compromise
        the security policies at the application layer itself. Downgrade attacks
        (arising from backward compatibility~\cite{poodle}) can be possible if
        an exploitable backward compatibility is provided by the user agents.
\item Already a set of security policies is getting expressed via domains (see
    \autoref{sec:pre:security_policies}). Hence, the Internet security may get
        benefited if domains can easily express and manage more security
        policies in future.
\end{enumerate}

For a better expressiveness of domain level security policies, we propose a
metapolicy framework through which domains can specify and manage a
comprehensive set of their security policies. The proposed framework leverages
the DNS infrastructure for publishing and accessing metapolicies, and the trust
infrastructures of TLS or DNSSEC to provide the necessary layer of security.

\section{Background}
\label{sec:pre}
\subsection{Domain Name System}
Domain Name System (DNS)~\cite{mockapetris1983domain} is a decentralized and
hierarchical system which stores information about domains. Different types of
information are stored in different resource records.  Some of the DNS resource
record types include \texttt{A} record that points a domain to an IPv4 address,
\texttt{CNAME} record that points one domain to another domain, \texttt{TXT}
record for storing human-readable textual information, or \texttt{MX} records
for point to domain's mail exchangers.  DNS is mostly known for resolution of
domain names to IP addresses, however currently, the DNS is getting utilized
for storage of email policies, information on domain certificates, and other
domain related information.  Publishing policies over DNS has an inherent
benefit. As most of the times a DNS resolution precedes the communication with a
domain, it is easy for the initiating party to fetch security policies prior to
the connection.  This also removes the need for communication with any other
party (only DNS servers are contacted). 

DNS Security Extensions (DNSSEC)~\cite{larson2005dns} is an extension of DNS
which provides security to the DNS records by adding cryptographic signatures on
top of it.  For each DNS zone a zone signing key (ZSK) pair and ZSK's private
key is used to sign the DNS records (the corresponding signatures are stored in
special \texttt{RRSIG} resource records).  The ZSK public key is stored in the
\texttt{DNSKEY} record. The \texttt{DNSKEY} record is also signed with the
private key of another key pair known as Key Signing Keys (KSK).  The chain of
trust is followed till the root. This addition of signature on top of DNS
records help in verifying the origin of the DNS records and in identifying if
the records have been tempered during the transit via a MITM attack.

\subsection{Transport Layer Security}
Transport Layer Security (TLS) is a key protocol that provides confidentiality
and data integrity on the Internet. The TLS handshake protocol is the initial
phase of the TLS, and it provides a way through which the clients and the
servers can verify each other identity via X.509 digital
certificates~\cite{housley2002internet} issued to them by trusted certification
authorities (CAs).

X.509 public-key infrastructure (PKI) certificates are issued to domains (such
as google.com) by trusted intermediate CAs (such as Google Internet Authority
G3) forming trust chains.  The certificate contains the details of the domain's
identity and the domain's TLS certificate's public key. The information in the
certificate is trusted as a trusted CA has signed it asserting its correctness.
X.509 certificates are either signed by other intermediate CAs or the root CA
and then a root CA (such as GlobalSign) may have a self-signed X.509 certificate
that is stored by clients.  The chain of trust can be verified till root CA to
identify if the certificate issued to the domain is valid.  Usually, only
servers have their certificates (i.e., clients' identities are not verified by
servers).

As communicating parties can verify their identifies, the TLS handshake protocol
allows them to securely exchange secret session keys. The session key is then
used for the encryption of data over a communication session between the clients
and the servers.

\subsection{Security Policies}
\label{sec:pre:security_policies}
Email and the TLS PKI are two key areas in which domains are currently
expressing their security policies. Email policies are one of the oldest
policies that rely upon the DNS infrastructure. 

The Sender Policy Framework (SPF)~\cite{rfc7208} helps the receiving email
server to identify whether the host from which the email has been originated is
an authorized entity to send an email to the domain's owner. Spam and phishing emails
can be filtered using this email policy. To deploy this policy the domain needs
to add a \texttt{TXT} record in its DNS zone file, specifying authorized
addresses (that can send emails on behalf of the domain).

DomainKeys Identified Mail (DKIM)~\cite{rfc6376} helps in verifying the
authenticity of a given email. A domain supporting DKIM digitally signs the
outgoing emails using a private key. The domain publishes the corresponding
public key in the DKIM-specific DNS \texttt{TXT} records. A receiving email
server accesses the public key from the DNS records of the email's originating
domain. This public key is used to verify the digital signature of the email.
DKIM aims to ensure that the email has not been modified in the transit and is
signed by the correct outbound email server authorized to send email for that
domain.

Domain Message Authentication Reporting and Conformance (DMARC)~\cite{rfc7489}
is a policy system that allows domain owners to specify whether SPF or DKIM or
both should be used while sending the emails for that domain and what the
receiving email servers should do in the case of policy failures.

DNS-based Authentication of Named Entities (DANE)~\cite{rfc6698} is a TLS PKI
policy system that  provides a way to authenticate TLS entities without a CA.
DANE introduces new \texttt{TLSA} records, that  are published over DNS and
signed are via DNSSEC. TLSA records provide domains a way through which they can
specify which CAs can issue a valid TLS certificate for a domain and which TLS
certificate to use for a specific service. If a browser supporting DANE get a
TLS certificate for a domain which is not from the domain specified CA list,
then it can display a warning to the user mentioning that the connection with
the domain is insecure.  

Certification Authority Authorization (CAA)~\cite{rfc6844} provides a mechanism
by which domains can specify (over DNS) which CAs can issue certificates
for them and their subdomains.  It is required for a CA to retrieve a CAA record
for a particular domain and follow the rules and restrictions before issuing a
certificate for that domain. 

Some policies are defined using HTTP headers, instead of employing the DNS
infrastructure.  For instance, HTTP Strict Transport Security
(HSTS)~\cite{rfc6797} allows web operators to mandate access to their websites
on HTTPS connections. Whenever a browser accesses a website for the very first
time the website replies back with an HSTS header that specifies that the
subsequent connections should be conducted over HTTPS.  The browser caches this
information and connects the website only via HTTPS even if the user types a URL
with HTTP specified.  Around 4.37\% of the domains enforce HSTS and there has
been an increase of around 69\% in its usage in Q2 2017~\cite{StatsHTTP}.

Similarly,  HTTP Public Key Pinning (HPKP)~\cite{rfc7469} is a policy mechanism
that allows domains to express their keys or keys of their CAs using HTTP
headers. Around 0.71\% of domains on an average are expected to have enforced
HPKP. There has been an increase of 42\% in the use of HPKP in Q2
2017~\cite{StatsHTTP}.  However, browser vendors decided to obsolete HPKP due to
operational issues~\cite{kranch2015upgrading}.
\medskip

The deployment of the presented policies was recently analyzed by Szalachowski and
Perrig~\cite{szalachowski2017short}, and Amann et al.~\cite{amann2017mission}.

\section{Requirements and Challenges}
\label{sec:overview}
In the current Internet, there is no comprehensive and secure framework through
which the security policies can be easily defined, managed, stored, and published
by domain owners. We identify a set of requirements that such a security
policy framework should follow to enhance the domain expressiveness on the
Internet. These include:

\begin{enumerate}
\item \textbf{Easy Management:} The new policy specification framework or
    protocol must make it easy for domains to specify, manage, and publish
        various security policies at one place with a sufficient level of
        security from known threats.
\item \textbf{Security:} The protocol must provide security for policies, i.e.,
    policy agents can verify their authenticity (i.e., that a given policy was
        indeed produced by the corresponding domain).
\item \textbf{Deployability:} The protocol must be easy to deploy, manage, and
    use. Moreover, policies should be disseminated and secured using the
        existing infrastructures to minimize operational and deployment costs.
\item \textbf{Recoverability:} The protocol should not end up in an
    unrecoverable state. It must provide suitable recovery mechanisms in the case
        of a policy misconfiguration.
\item \textbf{Adaptability:} The protocol must be adaptable in the sense that it
    can coexist with the currently deployed mechanisms without needing major
        changes.
\item \textbf{Availability:} Policies should be highly available and publicly
    accessible.
\end{enumerate}

\section{A High-level Overview}
To fulfill the above requirements we propose a comprehensive and secure
metapolicy framework for specification and management of domain security
policies. The framework allows the domains to specify, manage all the existing
domain-level security policies as a metapolicy. Metapolicies are published in
DNS and are secured using the existing TLS or DNSSEC PKI infrastructure. 

\begin{figure}
\begin{center}
  \includegraphics[width=0.9\linewidth]{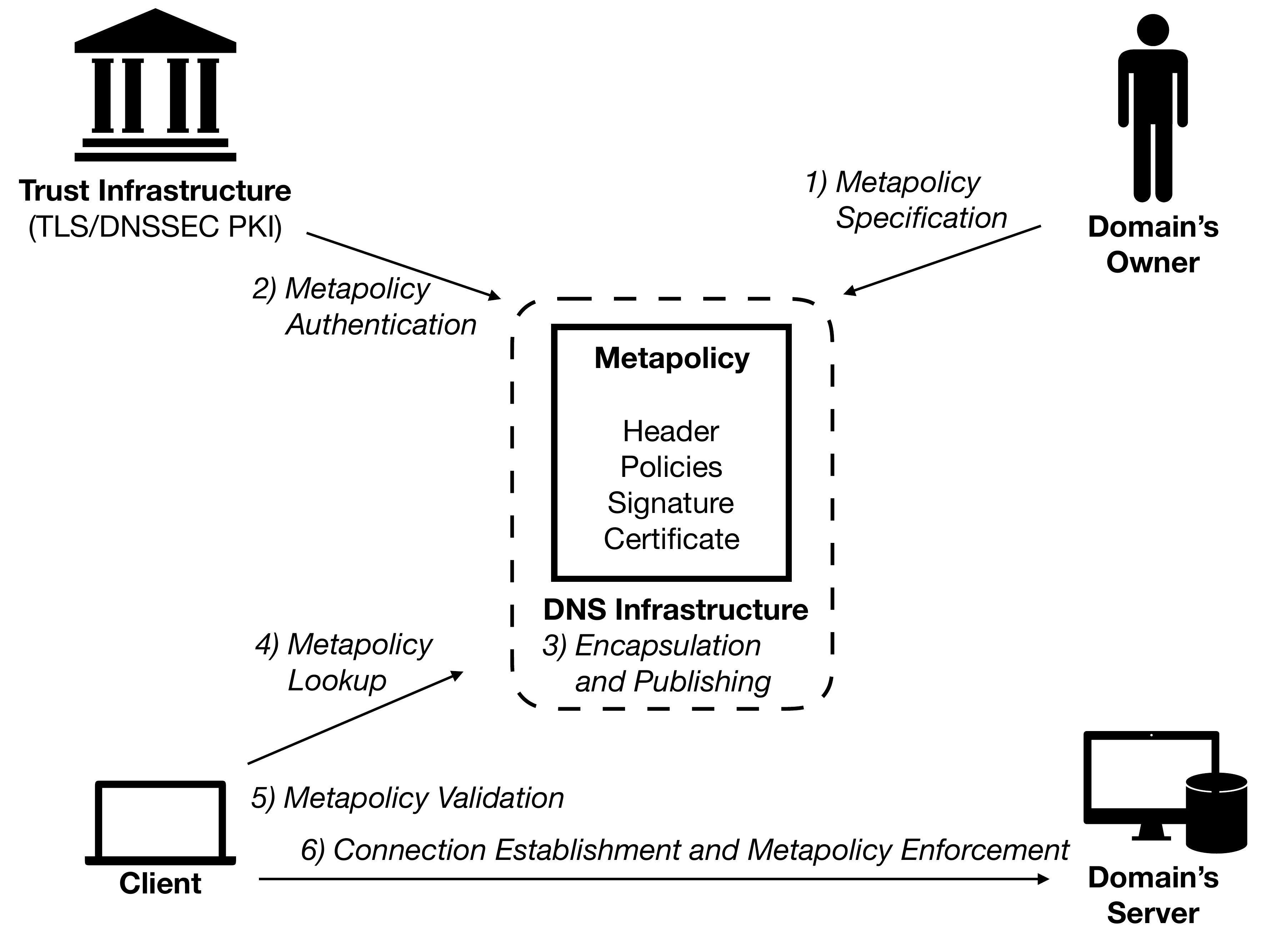}
    \caption{A high-level overview of the metapolicy framework.}
  \label{fig:boat1}
  \end{center}
\end{figure}
A high-level abstract overview of the proposed metapolicy framework is given in
\autoref{fig:boat1} and the sequential workflow is described as follows.

\begin{enumerate}
\item \textbf{Metapolicy Specification:} The domain-level policies are specified by
    the domain owners using the policy specification format of the metapolicy
        framework (details of the metapolicy format is given \autoref{sec:details}).
\item \textbf{Metapolicy Authentication:} The metapolicy is then signed using the
    domain's X.509 certificate private key or DNSSEC key.  Since the domain's
        TLS certificate (or the DNSSEC key) can be verified, the  domain binding
        with the metapolicy can be verified too.
\item \textbf{Encapsulation and Publishing:} Finally the signed metapolicy gets
    published in the DNS. To this end, the metapolicy has to be encoded as
        resource records. Publishing metapolicy in the DNS decreases the
        infrastructure cost and latency.
\item \textbf{Metapolicy Lookup:} Policies can then be queried by policy agents
    whenever a domain is going to be visited by a user (i.e., when a DNS
        resolution for a domain takes place).
\item \textbf{Metapolicy Validation:} The metapolicy's signature is verified
    using the domain's TLS certificate public key or the DNSSEC public
        key.  All information required to validate the metapolicy is published
        as its part.
\item \textbf{Connection Establishment and Metapolicy Enforcement:} Once the
    metapolicy is verified the content of the metapolicy (individual security
        policies) are extracted and the specifications are enforced by the
        policy agents during the access to the domain's services and resources. 
\end{enumerate}

\section{Details of the Framework}
\label{sec:details}
In the proposed framework all the domain security policies are included within a
single metapolicy. Every metapolicy consists of:

\begin{itemize}
    \item \texttt{Header}: This section contains metadata about the metapolicy.
    \item \texttt{Policies}: This section contains the actual content of the
        various security policies which are specified by the domain owner.
    \item \texttt{Signature}: This section contains a signature created using
        the domain's TLS certificate key or DNSSEC key over the metapolicy
        header and the policies section.
    \item \texttt{Certificate}: This section contains the domain's TLS
        certificate chain which is necessary to verify the authenticity of the
        created signature (i.e., whether the metapolicy was signed by the
        correct domain).  When the metapolicy is signed with the DNSSEC key this
        field is empty, as the DNSSEC key of the domain can be obtained through
        the \texttt{DNSKEY} record.
\end{itemize}
\medskip

The \texttt{Header} section contains the basic metadata about the metapolicy.
In particular, it includes the  following:
\begin{itemize}
    \item \texttt{Domain} name on which the metapolicy is applicable. This is
        stored as a string.
    \item \texttt{Version} number of the metapolicy. The version increments when the
        metapolicy changes and an update happens. The version is represented as
        an integer. For example, a value of 1 in \texttt{Version} will represent
        the first version of the metapolicy.
    \item \texttt{Valid From}, \texttt{Valid To} dates in the
        \textit{mm/dd/yyyy} format to specify the time period in which the
        metapolicy is considered as valid. Time is expressed in the UTC standard.
    \item \texttt{Parts} specify the number of DNS \texttt{TXT} records (see
        below) that needs to be downloaded to get the contents of the complete
        metapolicy. If the complete metapolicy can be wrapped up in 512 bytes the
        value of \texttt{Parts} is set to 1 else it will always be greater than
        1 and will correspond to the number of \texttt{TXT} records needed to
        store the complete metapolicy.  This field is required to encapsulate and
        decapsulate metapolicies over DNS protocol.
    \item \texttt{Subdomains} section lists the subdomains which will also
        follow the specified policies. Hence inheritance is provided as the
        information of whether the subdomains will follow the domain policies
        can also be specified in the metapolicy. This section can store
        subdomain names as a comma-separated list (it can be also a
        \textit{wildcard} domain).
\end{itemize}
\medskip

The \texttt{Policies} section contains the actual content of domain security
policies. Each policy has to specify these fields in the domain's metapolicy:
\begin{itemize}
    \item \texttt{ID} specifies a unique RFC number of a specific security policy.
    \item \texttt{Specification} section contains the actual content of a policy.
    \item \texttt{Fail} section instruct the clients about what they should do
        if a policy failure happens (an error in a policy specification or an
        error during its enforcement). The failing function can be either
        \texttt{hard}, \texttt{soft}, or \texttt{ignore} instructing the policy
        agent, that if a policy failure happens, the client should either
        immediately terminate the connection (\texttt{hard}), or soft-fail
        (\texttt{soft}) and show a warning to the user, or just \texttt{ignore}
        this policy failure and proceed normally.  Domains can also instruct
        clients to do error reporting to a set of email addresses in case of
        failures. 
\end{itemize}
\medskip

The \texttt{Signature} section stores the signature computed over the metapolicy
\texttt{Header} and \texttt{Policies} sections.  The key used for signing the
metapolicy corresponds to the private key(s) of the domain's TLS certificate
or domain's DNSSEC key. 

The last section of the metapolicy is the \texttt{Certificate} section that
stores the domain's X.509 certificate chain (i.e., domains certificate and
certificates of intermediate CAs).  This certificate chain is used by the
policy agents for validation of the domain's TLS certificate and the
signature of the metapolicy. The storage of all the certificates (required to
establish the chain of trust) in the domain's metapolicy avoids the extra
efforts of locating and downloading these certificates by the policy agents.
When the metapolicy is authenticated with the DNSSEC key this section is empty.

Finally, the complete metapolicy is published via DNS.  To do so, it has to be
encapsulated into DNS resource records. A natural resource record type to store
an arbitrary information is \texttt{TXT}.  However, as shown by Szalachowski and
Perrig~\cite{szalachowski2017short} to transmit resource records reliably, they
should not exceed 512 bytes.  Therefore, if the total size of the metapolicy
exceeds 512 bytes the metapolicy record is stored in parts up to 512 bytes each.
The first part is published at \texttt{\_metapolicy.<domain\_name>} and the
policy agents learn the number of parts by accessing the value of the
\texttt{Parts} field from the metapolicy header (located in the first part).
Other parts of a metapolicy are accessed by querying
\texttt{<part\_number>.\_metapolicy.<domain\_name>} (e.g.,
\texttt{2.\_metapolicy.fb.com}).

An example policy is shown in \autoref{fig:policy_example}.

\begin{figure}
\begin{Verbatim}[fontsize=\small,frame=single]
 Header:
    Domain: a.com
    Version: 1
    Valid From: 12/09/2016 UTC
    Valid To: 12/09/2018 UTC
    Parts: 1
    Subdomains: example.a.com, verbal.a.com
    
 Policies:
    Id: 7288
    Specification: v=spf1 a include:aspmx.googlemail.com ~all
    Fail: hard, report@a.com
    
    Id: 6376
    Specification:v=DKIM1; k=rsa; p=TAMAfMA0GCSqGSIb3DQLOGE...
    Fail: soft, report@a.com
    
 Signature: 9243152cd53fe3d1...
    
 Certificate: MIIEBDCCAuygAwIBAgIDAjJ... 
\end{Verbatim}
    \caption{An example of the metapolicy.}
    \label{fig:policy_example}
\end{figure}

\subsection{Metapolicy Lifetime}
\subsubsection{Creation}
A domain creates its metapolicy by specifying the security policies in the
format specified in \autoref{fig:policy_example}. The domain then digitally
signs the metapolicy with the private key(s) associated with its X.509 TLS
certificate or with its DNSSEC private key. Finally, the signed metapolicy is
published in the DNS as a series of \texttt{TXT} records.

\subsubsection{Querying and enforcing meta policies} 
Whenever a policy agent receives a request to connect to a domain it obtains the
domain's metapolicy (if not cached) from the DNS \texttt{TXT} records of that
domain.  However, if the metapolicy for a domain has already been cached by the
policy agent only the first DNS \texttt{TXT} record gets downloaded. The cached
metapolicy is utilized and the complete metapolicy from the DNS does not get
downloaded if the version of the metapolicy in the DNS is not higher than the
version of the cached metapolicy. 

Integrity and authenticity of the metapolicy content are guaranteed by the
digital signature. To validate a metapolicy the policy agent must verify the
\texttt{Signature} with the public key available from the domain's TLS
certificate or DNSSEC. The client must also verify the domain's TLS
certificate by validating the trust chain.  If the signature verification
succeeds the content of the specific security policies (identified by their
\texttt{ID}) are fetched and enforced by the policy agent. Policy failures are
handled and reported depending on the failing scenario specified
(\texttt{Fail}). 

A pseudocode that describes querying and enforcing of metapolicies is given in
\autoref{Algo2}.  

\begin{algorithm}
\caption{Querying and Enforcing Metapolicy}\label{Algo2}
\texttt{M\mysubscript{Domain}}:  Domain's metapolicy  \\
\texttt{S\mysubscript{Policy}}:  Metapolicy's signature \\
\texttt{DNS\mysubscript{TXT}}: DNS TXT records storing the domain's metapolicy. \\
\texttt{DNS\mysubscript{TXT Part 1}}: The first part of the metapolicy's DNS TXT record containing the metapolicy's header information. \\
\texttt{M\mysubscript{Domain}(Cache)}:  Client cached version of Domain's
    metapolicy specifications. \\
    \texttt{Cache}: Client's/Server's local storage to store the metapolicy. \\
    \texttt{Policy}: Stores the content of a security policy. \\
    \texttt{Return}: Stores the execution status of the metapolicy querying and enforcement operations. \\
\texttt{ID}: ID represents the RFC number of a specific security policy. \\
\texttt{\textit{Cached(X)}}: Checks if the metapolicy for domain X is
    cached in the client's local storage. \\
\texttt{\textit{FetchContent(X)}}: Fetches the content of a security policy identified by \texttt{ID} X. \\
\texttt{\textit{Verify(X)}}: Verify if the signature (\texttt{S\mysubscript{Policy}}) of the metapolicy (represented by X) is valid using the domain's TLS Certificate or DNSSEC key. \\
\texttt{\textit{Delete(X)}}: Deletes the contents of the metapolicy X from the client's cache. \\
\texttt{\textit{Enforce(X})}: Enforce the specifications of policy X and return the execution status as either success or failure (soft, hard, ignore). \\

\eIf{Cached(M\mysubscript{Domain})}{
	M\mysubscript{Domain} $\leftarrow$ DNS\mysubscript{TXT Part 1} \\
 	\eIf{M\mysubscript{Domain}(Cache) $\rightarrow$ Version \texttt{is equal to} M\mysubscript{Domain} $\rightarrow$ Version }
 	{
    Policy $\leftarrow$ FetchContent(\texttt{ID})     (\textit{From Cache})\\  
    Return $\leftarrow$ Enforce(Policy)\\
    }
    {
    Delete(M\mysubscript{Domain}(Cache))\\
    M\mysubscript{Domain} $\leftarrow$ DNS\mysubscript{TXT} \\
    \eIf{Verify(\texttt{S\mysubscript{Policy}}) == Success }
    {
    Policy $\leftarrow$ FetchContent(\texttt{ID})\\
    Cache $\leftarrow$ M\mysubscript{Domain} \\
    Return $\leftarrow$ Enforce(Policy)\\
    }
    {
    Return $\leftarrow$ \texttt{hard}\\
    }
    }
   }
{
    M\mysubscript{Domain} $\leftarrow$ DNS\mysubscript{TXT} \\
    \eIf{Verify(\texttt{S\mysubscript{Policy}}) == Success }
    {
    Policy $\leftarrow$ FetchContent(\texttt{ID})\\
    Cache $\leftarrow$ M\mysubscript{Domain} \\
    Return $\leftarrow$ Enforce(Policy)\\
    }
    {
    Return $\leftarrow$ \texttt{hard}\\
    }
}
\end{algorithm}

\subsubsection{Updates and Recovery}
An update happens when at least one of the metapolicy section needs to be
updated. The changes can be modifications of critical parameters (like adding or
removing of security policies); update of the \texttt{Valid From} and
\texttt{Valid To} field etc... In all cases, the metapolicy \texttt{Version} needs to
be updated and a new signature must be calculated and placed in the
\texttt{Signature} field of the metapolicy.

In the case when a cached metapolicy expires (i.e., the current date is greater than
\texttt{Valid To}) the policy agent will fetch a new metapolicy published by the
domain in the DNS. If by any chance the domain has not published a new
metapolicy (a metapolicy with higher \texttt{Version}) the policy agent will use
the cached metapolicy and report it to the domain.  Because the policy agent
queries the metapolicy header during each DNS query (i.e., each connection), it
will download the newly published metapolicy once it finds that the
\texttt{Version} number of the metapolicy in DNS is higher than that of the one
stored in its local cache.

If the private key of the domain's TLS certificate or DNSSEC
gets compromised or lost the last metapolicy published by the domain will still
remain valid. This is because the policy agents can still verify the metapolicy
using the domain's public key which will hold true until the TLS certificate
corresponding to the compromised key gets revoked or a new DNSSEC key pair is
generated and published. The certificate revocation does not affect the
metapolicy framework because the policy agents who have already cached an old
metapolicy will not be verifying the chain of trust again and whenever they 
find a higher version of metapolicy published in the DNS they will use the new
chain of trust to validate the domain's new TLS certificate or DNSSEC key which
is used to sign the metapolicy. Also, the metapolicy framework does not get
affected when some of the intermediate CAs (in the domain's TLS certificate
chain of trust) go out of business for the same reason. However, whenever a new
TLS certificate is introduced the domain must remove the old certificate from
the \texttt{Certificate} section and add the new certificate belonging to the
new chain of trust. If with that change a domain's private/public keypair was
changed, the domain must also update the old signature in the \texttt{Signature}
section.

\section{Analysis}
\label{sec:analysis}
\subsection{Security Analysis}
We assume that the first connection to the DNS is not under attack because if
that is the case then a MITM adversary could just censor all subsequent
communication and clients would never reach a metapolicy.  We also assume that
the user's system and the policy agent are trusted and that the system is free
from host-based malware. Study of the effects of malware on the security of the
proposal is currently out of the scope of the current research work. 

With the above assumptions the metapolicy framework can be compromised when:
(1) the policy agents or user does a wrong decision in case of policy failures,
or (2) when the key used to sign the metapolicy gets compromised or used PKI is
compromised.

For the first case, as all the information resides within the metapolicy and is
specified by the domain owners; the policy agents or the users are not involved
in decision making during policy failures. Hence attacks arising from user's bad
decision making or from provisions of backward compatibility cannot happen if
the domain does not specify to take a user input or want the policy agents to
fallback during a policy failure. The possibility of downgrade attacks also gets
reduced with the use of our metapolicy framework because the policy agents can
cache the metapolicy records.

An adversary able to compromise a domain's private key, or able to obtain a
malicious certificate on behalf of the domain can create a malicious metapolicy.
In such a case, the domain owner can initiate the recovery mechanisms, revoking
the malicious public key and establishing a new metapolicy.

\subsection{Deployability} \label{deploy}
As the proposed scheme uses the TLS or DNSSEC key(s) for signing the metapolicy,
all the domains supporting DNSSEC or TLS can deploy the proposed metapolicy
framework. To find out how many domains can possibly deploy our scheme we
conducted an experiment over a dataset of 120K top websites received from the
Alexa top 1 million domains list~\cite{AlexaTop1M}. We used the
\texttt{tls-scan} library~\cite{TLSSCAN} to obtain these statistics.  From
our experiments, we identified that around \texttt{77.8\%} websites support TLS
and \texttt{2.6\%} of the websites supports DNSSEC. Hence, a large fraction of
websites can implement the metapolicy framework even today.

We also measured the percentage of domains which may get benefited via
metapolicy framework. To calculate the same we conducted an experiment to obtain
the number of websites that today implement a security policy that can be
expressed by our metapolicy framework. The \texttt{host} command of Linux was
used to fetch the  records of various email and TLS policies from DNS. The
outcomes of the experiment are given in \autoref{table:table1}.
The obtained results indicate that majority of domains (around 76.3\%) sets at
least one security policy today.

\begin{table}
\caption{Number of websites supporting various domain policies}\label{table:table1}
\begin{center}
\begin{tabular}{ |c |r |r|  }
\hline
 Policy & Supporting & Percentage  \\
 & Websites & \\
 \hline
 SPF & 68213 & 56.00\% \\  
 DKIM & 56704 & 46.60\% \\
 DMARC & 11973 & 9.80\% \\
 DNSSEC & 3217 & 2.60\% \\
 CAA & 1213 & 0.99\% \\
 DANE & 34 & 0.03\% \\
 \hline
\end{tabular}
\end{center}
\end{table}

\begin{table}
\caption{Number of domains supporting multiple security policies}\label{table:table2}
\begin{center}
\begin{tabular}{ |c | r |r | c| r| r|  }
\hline
 \# of Policies & \# of Domains & Percentage & \# of Policies & \# of Domains & Percentage \\
 \hline
 At least 1 & 92801 & 76.30\% & 1  & 53057 & 43.62\% \\  
 At least 2 & 39744 & 32.67\% & 2  & 31755 & 26.24\% \\   
 At least 3 & 7989 & 6.50\% & 3  & 7233 & 5.94\% \\
 At least 4 & 756 & 0.62\% &  4 & 699 & 0.57\%\\
 At least 5  & 57 & 0.05\% &  5 & 50 & 0.04\% \\
 At least 6  & 7 & 0.01\% & 6 & 7 & 0.01\% \\
 \hline

\end{tabular}
\end{center}
\end{table}

\subsection{Overheads}
\subsubsection{Metapolicy Size}
Size of a TLS certificate chain is a dominant factor in the overall size of a given
metapolicy.  To find out how big this overhead is we conducted an experiment.
During this experiment, we downloaded all certificate chains which are required
for domain's TLS certificate validation for a domain set. We used the
\texttt{openssl} tool for this purpose.  The experiment was performed on the
Alexa top 13k websites.  We found that the average size of the of a certificate
chain needed for a domain's TLS certificate validation is around 4.75 KB. Thus
on average, a metapolicy protected with a TLS certificate will have to contain
4.75 KB for a certificate chain.  (Note that policy agents do not have to store
certificates of validated policies.)

To calculate the size of an average metapolicy we did an analysis of the results
obtained in \autoref{deploy}.  As shown in \autoref{table:table2}, around
33\% of websites deploy at least two or more policies. With the results from
\autoref{table:table1} we can assume that on average policies implemented by
domains will be either a SPF, DKIM or DMARC policy. We used this analysis to
identify the size of an average metapolicy record. We created multiple
metapolicy records with these three policies specified in it and stored domain's
TLS certificate chain and a computed signature. We calculated the
average size of metapolicy to be around 5.4 KB. Thus, on average, a metapolicy
would require about 11 \texttt{TXT} records to be encoded.

\subsubsection{Latency}
Another overhead is the additional time needed for fetching a metapolicy. To
calculate this overhead we performed an experiment sending DNS queries to
calculate the time needed for fetching a single DNS \texttt{TXT} record. We
identified that accessing it takes around 20 ms on an average, on a system
having a network download speed of 13 Mbps. In the same setting obtaining
additional 10 records, even sequentially (what is the worst case), increases the
latency by 200 ms (for the records queried in parallel that should be around
only 20 ms).  Hence, the proposed metapolicy framework introduces an acceptable
overhead on top of a normal DNS query for a metapolicy. However, once the
metapolicy is cached only the first 512 bytes of the metapolicy (the first part)
gets downloaded by the policy agents.

\subsubsection{Computational overhead}
To identify the overheads of the certificate validation process (that will
happen when the metapolicy's signature will be verified at the client) we used
the \texttt{OpenSSL} library and the certificate chains obtained in the previous
experiment. In our tests, we identified that it takes 4 ms on an average for the
certificate chain and signature validation process. Hence the metapolicy
verification introduces an acceptable overhead to a standard connection
establishment.

\section{Implementation}
\label{sec:implementation}
To implement a prototype of the proposed metapolicy framework, we used the Bind
open source DNS server implementation. We configured a Bind to serve as a
private DNS server.  It ran under Ubuntu 16.04 equipped with Intel (R) Core (TM)
i7-7600U CPU (2.8 GHz) with 8 GB of RAM. We created and published (in
\texttt{TXT} records) an example metapolicy.  We also prototyped a policy agent
able to fetch and process metapolicies.  Our experiments confirm the feasibility of
our framework and deployability even with currently existing tools and libraries.

\section{Related work}
\label{sec:related}
Despite important of the topic, there has been a little work in the area of
domain expressiveness over the Internet.  In particular, we are not aware of any
work which  directly fits into our line of research work described in this
paper.  
One example of domain expressiveness system is DMARC~\cite{rfc7489}. It is
is a policy system that allows domain owners to manage their email security
policies (SPF and DKIM, specifically). DMARC, similarly to our system, uses DNS
for publishing its policies. However, the scheme does not provide any security
and has limited functionality. 

Another related system is
PoliCert~\cite{Szalachowski:2014:PSF:2660267.2660355} which enhances the
security of the existing TLS PKI infrastructure by allowing domain owners to
decide and define policies that govern the usage of their TLS certificates. The
authors introduced the concept of subject certificate policies that provide
domains a way to specify trusted CAs, their update criteria, error handling and
private key loss mechanisms. To take care of a single CA compromise they
introduced the concept of multiple signature certificates that allows multiple
CAs to sign a certificate. PoliCert relies on verifiable public logs, thus it
needs to introduce a new infrastructure.

\section{Conclusions}
\label{sec:conclusions}
In this paper, we presented a metapolicy framework for enhancing the domain
expressiveness on the Internet. Our proposal provides domains a mechanism to
define and manage domain related security policies themselves. All the
metapolicies related to a domain and which the domains want to enforce can be
mentioned in a metapolicy which is signed by the domain's private key
corresponding to the domain's TLS certificate or DNSSEC key. The metapolicy is
published as a series of DNS \texttt{TXT} records in the domain's DNS zone.
Therefore, no new infrastructure is required, and our scheme can be deployed
today.

The framework makes it easy for domains to manage the policy themselves. It
also reduces the chances of a downgrade attack due to incorrect choices which
can be made by a user or its user agent, because a fail-over mechanism as
specified in the metapolicy has to be followed and neither the software or the
user decides the fate of a policy failure. It also provides a simple way of
management and specification of policies including the HTTPS related security
policies likes HSTS or HPKP or Email related security policies including SPF,
Sender ID, DMARC, DKIM or other security policies including the DANE or CAA.  In
future, we believe that more security policies can be expressed by domains
through our proposed metapolicy framework.

\section*{Acknowledgment}
We thank the anonymous reviewers whose feedback helped to improve the paper.
This work is supported by SUTD SRG ISTD 2017 128 grant.

\newpage

\bibliographystyle{plain}
\bibliography{ref,rfc}
\end{document}